\input{aipcheck.tex}

\newcommand\sps{\space\space\space\space}
\documentclass[
  ]
  {aipproc}

\newcommand\dz{D\O \space}

\layoutstyle{6x9}

\SetInternalRegister\hbadness{8000} 

\newcommand\doingARLO[2][]{%
  \ifx\mmref\undefined #1\else #2\fi
}

\begin{document}

\title 
      [dzero search]
      {\dz Search for the Higgs Boson in Multijet Events}


\author{Alex Melnitchouk for the \dz Collaboration}{
  address={Department of Physics and Astronomy,\\ 
           The University of Mississippi,\\ 
           108 Lewis Hall, P.O. Box 1848\\
           University, MS 38677-1848, USA.}
}


\copyrightyear  {2005}

\begin{abstract}
We present two searches for the Higgs boson in $\sqrt{\mbox{s}} = 1.96 \: \mbox{TeV} \:\mbox{p}\bar{\mbox{p}}$ collisions
using data taken with the \dz detector during Run II of the Fermilab
Tevatron collider. The first study is a search for neutral Higgs bosons 
produced in association with bottom quarks. 
The cross sections for these processes are enhanced in many extensions
of the standard model (SM), such as in its minimal supersymmetric
extension at large tan$\beta$. The results of our analysis agree with
expectations from the SM, and we use our measurements to set upper limits
on the production of neutral Higgs bosons in the mass range of 90 to 150 GeV. 

The second study is a search for the standard model Higgs boson produced
in association with the Z boson. 
We study the $\mbox{p}\bar{\mbox{p}}\to \mbox{ZH} \to \nu \bar{\nu} \mbox{b} \bar{\mbox{b}}$ channel, 
which is one of the most sensitive ways to search for light Higgs bosons 
at the Tevatron.
We select multijet events with large imbalance in transverse
momentum and two b-tagged jets. 
Then we search for a peak in invariant mass distribution of two b-tagged jets. 
After subtracting the backgrounds, we set the 95\% C.L. upper limits on the
$\sigma(\mbox{p}\bar{\mbox{p}}\to \mbox{ZH} \to \nu \bar{\nu}) \times \mbox{BR(H}\to \mbox{b} \bar{\mbox{b}})$
for Higgs masses between 105 and 135 GeV.
\end{abstract}

\date{\today}

\maketitle

\section{Introduction}

The Higgs boson, which is hypothesized to be responsible for electroweak symmetry breaking,
is the only particle in the SM that has not been directly observed.
Higgs boson also appears in the extensions of the
SM, such as supersymmetry (SUSY). 
In the minimal supersymmetric extension of the SM (MSSM) ~\cite{mssm},
there are two Higgs fields and five physical Higgs bosons, three of which are neutral.
Tevatron experiments can be sensitive to both SM Higgs boson and MSSM neutral Higgs bosons.

In both searches presented in this paper  ~\cite{hz,bbh}, final states contain b-jets.
Higgs signal would appear as a "bump" in the di-b-jet invariant distribution.
Therefore, good understanding of the calorimeter response and b-tagging are the main ingredients
of both analyses.

\section{Standard Model Higgs Search in Multijet Events}

We use data collected by the \dz detector ~\cite{Dzero} 
between March  2003 and June 2004, corresponding
to an integrated luminosity of about 261 pb$^{-1}$.
We select events with two acoplanar b-jets with transverse energy above 20 GeV.
We require that missing transverse energy in the event is above 25 GeV.
In addition, we use kinematic variables based on vector sum of transverse momenta of jets and tracks.

We distinguish between two types of background, such as physics background and instrumental background.
Physics backgrounds are mainly due to misidentified b-jets in Z+jets and W+jets processes with final state neutrinos that escape the detector.
Instrumental backgrounds are multijet events in which energy of jets is mismeasured.
Physics backgrounds are estimated using using PYTHIA, COMPHEP, and ALPGEN Monte Carlo event generators.
Monte Carlo samples were processed through the \dz detector simulation and reconstruction software.
Instrumental background is estimated from data.

\begin{figure}[zhlimits]
  \includegraphics[height=.35\textheight]{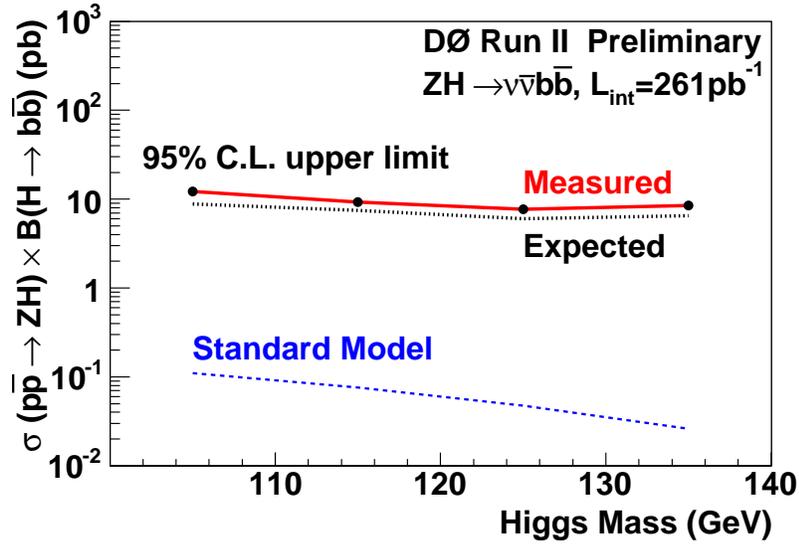}
\caption{95 \% C.L. upper limits on the cross section for ZH production 
        times the branching fraction for $\mbox{H} \to \mbox{b} \bar{\mbox{b}}$}
\label{fig:zh}
\end{figure}

No excess of events is observed.
We derive upper limits on 
$\sigma(\mbox{p}\bar{\mbox{p}}\to \mbox{ZH}) \times \mbox{BR(H}\to \mbox{b} \bar{\mbox{b}})$
for four Higgs mass points in the range between 105 and 135 GeV.
The results are plotted in Figure ~\ref{fig:zh}.

\section{Supersymmetric Higgs Search in Multijet Events}

We require three jets above 15 GeV at the highest trigger level.
In the offline we require at least three b-tagged jets.
Both the $\mbox{p}_{\mbox{T}}$ thresholds and pseudorapidity range selections
are optimized for Higgs mass and number of required jets.
Signal rates and kinematics are normalized to NLO calculations.

Major source of background is SM multijet production.
There are two main categories of multijet background.
One contains genuine heavy-flavor jets, while the other has only light-quark
or gluon jets that are mistakenly tagged as b-quark jets, 
or correspond to gluons that branch into nearly collinear b$\bar{\mbox{b}}$
pairs.
Background shape is determined from double b-tagged data by applying the
tag rate function to non-b-tagged jets. Then the background is normalized
to the data outside the signal region.

\begin{figure}
\hspace{-0.35cm}
\includegraphics[height=.33\textheight]{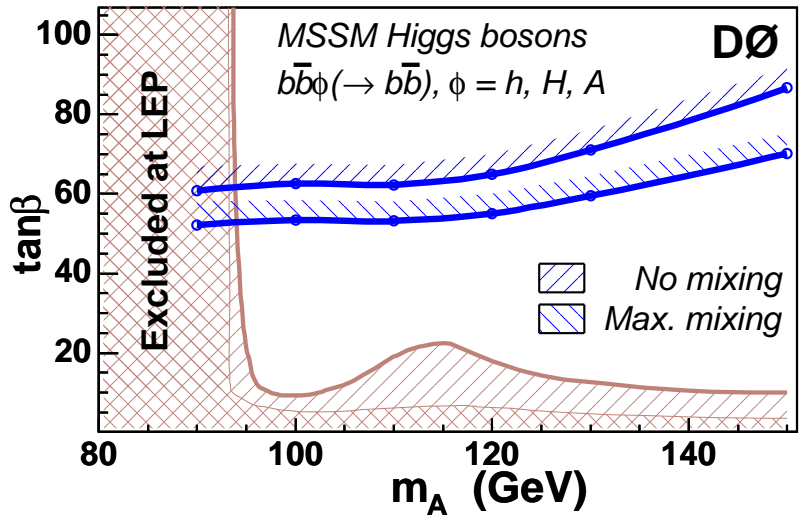}
\caption{The 95\% C.L. upper limit on tan$\beta$ as a function of $m_{A}$
for two scenarios of MSSM, ``no mixing" and "maximal mixing". Also shown
are the limits obtained by the LEP experiments for the same two
scenarios of the MSSM \protect\cite{lep}}
\label{fig:bbh}
\end{figure}

No excess of events is observed and
upper limits on neutral Higgs production cross-section are derived.
The limit is shown as a function of Higgs mass and tan$\beta$
in Figure ~\ref{fig:bbh}.

\begin{theacknowledgments}
%
We thank the staffs at Fermilab and collaborating institutions, 
and acknowledge support from the 
DOE and NSF (USA);
CEA and CNRS/IN2P3 (France);
FASI, Rosatom and RFBR (Russia);
CAPES, CNPq, FAPERJ, FAPESP and FUNDUNESP (Brazil);
DAE and DST (India);
Colciencias (Colombia);
CONACyT (Mexico);
KRF and KOSEF (Korea);
CONICET and UBACyT (Argentina);
FOM (The Netherlands);
PPARC (United Kingdom);
MSMT (Czech Republic);
CRC Program, CFI, NSERC and WestGrid Project (Canada);
BMBF and DFG (Germany);
SFI (Ireland);
Research Corporation,
Alexander von Humboldt Foundation,
and the Marie Curie Program.

\end{theacknowledgments}


\doingARLO[\bibliographystyle{aipproc}]
          {\ifthenelse{\equal{\AIPcitestyleselect}{num}}
             {\bibliographystyle{arlonum}}
             {\bibliographystyle{arlobib}}
          }

\end{document}